**Structure and dynamics of a pinned vortex liquid in superconducting *a*-Re$_x$Zr (x ≈ 6) thin film**

Rishabh Duhan[1], Subhamita Sengupta[1], Ruchi Tomar[1], Somak Basistha[1], Vivas Bagwe[1], Chandan Dasgupta[2] and Pratap Raychaudhuri[1*]

[1] *Tata Institute of Fundamental Research, Homi Bhabha Road, Colaba, Mumbai 400005, India.*
[2] *Department of Physics, Indian Institute of Science, Bangalore 560012, India.*

We report the formation of a pinned vortex liquid spanning a very large region of the magnetic field-temperature parameter space in a 5 nm thick amorphous superconducting Re$_x$Zr (x ≈ 6) (*a*-ReZr) thin film, using a combination of low-temperature scanning tunnelling spectroscopic (STS) imaging and magnetotransport measurements. The nature of the vortex liquid differs significantly from a regular liquid. Analysing series of STS images captured as a function of time, we observe that the interplay of pinning and intervortex interactions produces a very inhomogeneous state, where some vortices remain static, whereas others move forming a percolating network along which vortices are mobile. With increase in temperature or magnetic field this network becomes denser eventually encompassing all vortices. Our results provide key insight on the nature of a pinned vortex liquid and some of the peculiarities in the transport properties of ultrathin superconducting films.

---

[*] E-mail: pratap@tifr.res.in

In recent years there has been renewed interest in the mixed state in thin superconducting films, owing to the observation of a variety of vortex fluid states[1,2,3,4] in these materials. On general grounds, it is expected that in a clean superconductor the crystalline vortex lattice can melt at a characteristic temperature/magnetic field, producing a vortex liquid (VL) state. So far, VL states in bulk single crystals have been unambiguously detected only in layered High-$T_c$ cuprates[5,6,7,8] at large operating temperatures. On the other hand, for conventional superconductors, the vortex lattice is either seen to retain its crystalline structure till very close to the upper critical field ($H_{c2}$) line[9,10] (in extremely clean crystals) or is driven into a disordered vortex-glass above the order to disorder transition[11,12,13]. The situation is different in thin films, where reduced dimensionality renders the vortex lattice more susceptible to thermal (and quantum[14,15]) fluctuations. Indeed, in weakly pinned superconducting thin films, two-dimensional (2D) VL states with different long range and short-range correlations, such as isotropic, hexatic and smectic, have been identified using real space imaging of the vortex state[2,3,4].

When a current is passed in the mixed state of a superconductor, each vortex experiences a Lorentz force, $F_L = J\Phi_0 t$ (where $J$ is the current density, $\Phi_0 = h/2e$ is the flux quantum and $t$ is the thickness of the sample) which can cause the vortices to move and give rise to dissipation. However, defects that are inevitably present in the solid can pin the vortices, inhibiting this motion. When vortices form a crystalline solid, a finite density of weak pinning centres can collectively pin all the vortices, such that the dissipation happens only above a critical current[16,17], $I_c$, where the $F_L$ overcomes the pinning force. The effect of defect pinning in a VL is still an open question. In a VL vortices do not form a rigid lattice and one would expect that a finite density of pins cannot pin all vortices, except for the extreme situation where each vortex is individually pinned. However, it has been theoretically predicted that in 2D, such a vortex glass is unstable[18,19,20] except at $T = 0$. On the other hand, various inhomogeneous liquid states[21,22] such as interstitial liquid or vortex slush[23,24,25] have been proposed, where the motion of vortices is dictated by a combination of defect pinning and interaction with nearby vortices. Understanding the structure and dynamics of the pinned VL is not only important to realise the application potential of these materials, but also connects to the fundamental problem of pinned liquid that is relevant for a variety of systems; e.g. colloids[26], skyrmion lattices[27], liquid-crystal molecules or lipids in membranes[28].

Scanning tunnelling spectroscopy (STS) imaging has emerged as a powerful tool to obtain structural information of the vortex lattice[29]. Here, a fine conducting tip of a low temperature scanning tunnelling microscope[30] (STM) is brought within the tunnelling range of a superconductor, and the local density of states (LDOS) is measured from spatially resolved tunnelling conductance map, $G(V, \boldsymbol{r}) = \frac{dI(r)}{dV}\big|_V$ ($\boldsymbol{r}$ is the local coordinate on the surface), while rastering the tip over the sample surface. Since both superconducting energy gap and the coherence peak in the LDOS is suppressed at the normal core of a vortex, each vortex manifests as a local minimum[13] in $G(V, \boldsymbol{r})$ when the bias voltage is kept close to the coherence peak. This technique has been extensively used to image crystalline and glassy vortex states. However, application of STS in VL is trickier. Since STS imaging is a slow measurement, imaging of moving vortices has mostly been restricted to situations where the motion is very slow[1,3,4,31,32,33]. However, when the movement of vortices is faster than the characteristic time of data acquisition at every pixel (few milliseconds), rapid fluctuation due to moving vortices will get integrated out, thus blurring the boundary between the vortex core and the gapped superconducting region outside[14]. The measured $G(V, \boldsymbol{r})$ at any point will thus be an average of the contribution both from inside and outside the vortex core, and its value will depend on how frequently a vortex came below the tip. Consequently $G(V, \boldsymbol{r})$ can be considered as a metric of the probability density of vortices, where its lowest value would correspond to a static vortex located at point $\boldsymbol{r}$, and the highest value would be for a location, $\boldsymbol{r}$, that is far from any vortex. For a homogeneous vortex liquid where the probability of finding a vortex is the same everywhere, $G(V, \boldsymbol{r})$ would be spatially uniform and featureless, and its value will be in between these two extremes. On the other hand, if there is a set of preferred points, $\boldsymbol{r}_i$, where vortices spend longer time (as expected for a pinned liquid), $G(V, \boldsymbol{r})$ would display local minima at $\boldsymbol{r}_i$ but the value at the minima, $G^{lmin}(\boldsymbol{r}_i)$, will depend on the fraction of time a vortex spends there. In this interpretation, $G(V, \boldsymbol{r})$ provides a measure[34] of the time-averaged LDOS of vortices at point $\boldsymbol{r}$. In this paper, we use this information from STS images to investigate the structure and dynamics of pinned vortex liquid in a thin superconducting film.

Our sample consists of a 5 nm thick amorphous Re$_x$Zr, x ≈ 6 (*a-ReZr*) superconducting film ( $T_c$ ~ 4.65 K ) grown on oxidised Si substrate using pulsed laser deposition[35]. The coherence length and penetration depth are, $\xi \sim 7\ nm$ and $\lambda(T \to 0) \sim 800\ nm$ respectively. STM topographic image shows a very smooth surface with very low density of particulates[36]. In this sample, pinning is not deliberately introduced, but rather arises naturally from substrate imperfections, and increases as the film thickness is reduced. For STS measurements the sample was transferred directly from the deposition chamber into the STM using an ultra-high vacuum suitcase without exposure to air. For transport measurements, the same sample was subsequently covered with a 1.5 nm thick Ge capping layer to prevent oxidation, and then patterned into a Hall bar using argon ion beam milling. Resistance ($R$) as a function of temperature ($T$) and magnetic field ($H$) was measured in conventional 4-probe geometry in a $^3$He cryostat.

We begin with magnetotransport properties. Fig. 1(a) shows *R-T* in various magnetic fields ($H$) measured using a very small applied current, $I$ = 500 nA, which is more than 2 orders of magnitude smaller than $I_c$ obtained by linearly extrapolating back the slope of the flux flow region[37] of the current-voltage (*I-V*) characteristics at low temperatures (Fig. 1(c)). The transport is thus in thermally activated flux flow (TAFF) regime, where the vortex motion happens through thermally activated jumps over pinning barriers. In a VL for $I \ll I_c$, TAFF is predicted to give a temperature dependent linear electrical resistance[21]. In our sample, at 410 mK we can resolve this linear region above 40 kOe (Fig. 1(d)). We observe that the *R-T* curves follow a simple activated behaviour, $R = R_0 e^{-U(H)/kT}$, from few Kelvin below $T_c$ down to the temperature where data hits the noise floor (Fig. 1(b)). This activated behaviour is a hallmark of the putative pinned VL[21]. We do not observe any transition to a vortex glass, which would have resulted in an abrupt (faster than exponential) decrease in resistance at a characteristic temperature.

We now concentrate on the STS data. We first focus on the data taken at 410 mK, 10 kOe. Except otherwise stated magnetic field was always applied after cooling the sample to the base temperature in zero field. We acquired 10 successive $G(V, \boldsymbol{r})$ maps over an area containing approximately 120 vortices, where each map takes approximately 12 minutes to acquire. In Fig. 2 (a)-(c) we show the normalised conductance maps, $G_N(V, \boldsymbol{r})$, for the 1st, 4th, 7th image. The conductance scale is normalised and stretched such that the lowest and highest values of $G_N(V, \boldsymbol{r})$ across all 10 images correspond to -1 and 0 respectively. First, we concentrate on individual images. $G_N(V, \boldsymbol{r})$ maps show the presence of local minima ($G_N^{lmin}(V)$) where the vortices are preferentially located, forming a disordered lattice with short range hexagonal coordination. However, the values of $G_N(V, \boldsymbol{r})$ at the minima as well as in between two adjacent minima have wide variations. Only some of these local minima are deep, i.e. $G_N^{lmin}(V) < -0.8$, and are separated from adjacent minima with barriers where $G_N(V, \boldsymbol{r})$ is high. These minima appear at around the same location in every image. Here the vortices are localised and undergo only small displacement around their mean positions. On the other hand, we see the presence of several shallower minima corresponding to sites where the vortices are partially localised. All these sites contain at least one neighbouring shallow minimum to which the vortex can hop; since this necessarily means that there is also a finite probability to find the vortex anywhere between these two minima, the maximum along the line joining them is also suppressed. To elucidate this more clearly, we connect all adjacent minima for which $G_N^{lmin}(V) > -0.8$ and where along the line joining the two minima, $G_N(V) < -0.4$ using thin blue lines. The connected points define the motion paths along which the vortices hop from site to site within the timescale of acquisition of a single image. We also notice that the positions of some of these shallow minima change from one image to another. The likely reason for this is that some vortices do not undergo a "complete" hop from one minimum to another, but instead get trapped in a pinning centre midway, so that the corresponding $G_N(V, \boldsymbol{r})$ minima appear at a different location in the next image. In order to understand the correlation between the structure of the lattice formed by the potential minima and dynamics of vortices, in each image first we Delaunay triangulate the positions of the minima (Fig. 2(d)) to uniquely define the nearest neighbours for each point, and calculate 3 metrics.

We quantify the degree of localisation of vortices by calculating the average value of all the conductance minima $\langle G_N^{lmin}(V) \rangle$ as well as the fraction of minima that are connected to the motion paths ($f$). On the other hand, the structural distortion is quantified using two metrics: (i) $\sigma_l = \left( \frac{1}{N_l} \sum_i (l_i - a_H)^2 \right)^{1/2}$, where $l_i$ is the length of the $i$-th bond, $a_H = \left( \frac{\Phi_0}{2\pi H} \right)^{1/2}$ is the expected lattice constant for a hexagonal vortex lattice and $N_l$ is the total number of bonds in the image, and (ii) $\sigma_\theta = \left( \frac{1}{N_\theta} \sum_j \left( \theta_j - 60° \right)^2 \right)^{1/2}$, where $\theta_j$-s are the angles formed by two adjacent bonds starting from the same point, and $N_\theta$ is the total number of such angles. For a perfect hexagonal lattice, $\sigma_l = \sigma_\theta = 0$. In Fig. 2(e) we plot $\langle G_N^{lmin}(V) \rangle$, $\sigma_l, \sigma_\theta$ and $f$ for each image. We observe a clearly discernible correlation: When the $\sigma_l$ and $\sigma_\theta$ are low implying that the lattice is less distorted, $\langle G_N^{lmin}(V) \rangle$ and $f$ is also low implying that the vortices on the average are more localised and vice-versa. Therefore, we conjecture that the vortex motion is associated with a successive build-up of local strain and strain relaxation that happens over timescale of minutes. However, the actual motion of vortices is likely to be much faster, which we cannot determine from our experiments.

We now investigate the motion over longer time scales. In Fig. 2(f) we plot all the motions paths obtained from all 10 images in a single frame (blue lines), which form an interconnected network. In the same figure we also show the positions of minima obtained from 10 successive images using filled circular dots of different colour for each image. To capture the network arising from motions that involve an incomplete hop we also connect any two points on this image that are separated by a distance, $d < 0.5a_H$ (thick red lines)[36], eliminating any isolated segment that is shorter than $0.5a_H$ in length, since those correspond to meandering motion of a vortex about its mean position. The two networks together capture the global motion paths of vortices. For most parts, the blue and red networks overlap. However, it is noteworthy that minima from two successive images do not necessarily appear as adjacent points on this global motion path, implying that the temporal evolution of the minima could be faster than the time interval between imaging the same location in two successive images. To find out how robust this network is we repeated the measurement over the same area by heating the sample above $T_c$ and cooling it back in the field-cooled protocol (Fig. 2(g)).

Despite the large thermal cycling and preparing the vortex state using a different protocol we observe that global motion paths are similar showing that the global motion paths of vortices are primarily dictated by underlying defect pinning.

We can now look at the evolution of these global paths with field and temperature. First, we concentrate on the magnetic field evolution. Fig. 3(a) shows $R$-$H$ measured at 410 mK. In Figs. 3(b)-(f) we show the gradual evolution of the network of global motion paths with increasing field constructed from 10 images in the same way. With increase in field the network become denser, eventually, connecting almost all local minima across the 10 images. To quantify this, we also plot in Fig. 3(a) the fraction of all minima in all 10 images, $f_G$, that are connected to at least one motion path. $f_G$ increases smoothly with magnetic field. It is interesting to note that even though $R$ falls below our resolution limit of 5 m$\Omega$ below 40 kOe, down to 1 kOe we do not observe[36] a situation where the local minima in $G_N(V, \boldsymbol{r})$ appear at the same location in all 10 images, as would have been expected for a vortex glass. We conclude that the vortices remain in a liquid state even when their mobility is too small to produce a detectable resistance in transport measurements.

The temperature evolution of global motion paths at 10 kOe are shown in Figs. 4(a)-(e). All images are acquired on the same area by correcting any thermal drift using topographic markers. Here too, $f_G$ increases with temperature (Fig. 4(f)). Here, the additional interesting observation is that the number of vortices connected to the blue network ($N_B$) initially increases more rapidly than those connected to the red network ($N_R$), giving rise to a peak in the ratio $\left(\frac{N_B}{N_R}\right)$ (inset Fig. 4(f)). This can be understood as follows. The red network arises from a change of position of a $G_N(V, \boldsymbol{r})$ minimum which happens whenever the vortex gets trapped in a pinning center between two local minima in $G_N(V, \boldsymbol{r})$. As the temperature is increased these trapping events become rarer and the vortex movement predominantly involves hops between two $G_N(V, \boldsymbol{r})$ minima that leave their positions unchanged. (The opposite happens at very low fields, i.e. 1 kOe, and low temperatures where the minima are far apart and a complete hop from one minimum to another becomes extremely rare[36].) Above the peak temperature the red network also increases and $\left(\frac{N_B}{N_R}\right)$ approaches unity. We believe that this decrease marks the gradual cross-over towards a homogeneous unpinned liquid as theoretically predicted[21,38]. Above 2.8 K we cannot resolve minima in $G_N(V, r)$ anymore. Similar temperature evolution is also observed[36] at 20 kOe.

In summary, analysing STS conductance maps, we show the existence of an inhomogeneous VL in a conventional superconducting film formed under the combined influence of intervortex interactions and defect pinning. Our study suggests that the percolative nature of transport in this inhomogeneous landscape should be factored in to understand some of the unusual *I-V* characteristics observed in thin superconducting thin films, such as the onset of non-linear creep at very low currents at low temperatures[20,39,36] or the plateauing of resistance at a value lower than the flux-flow resistance (vortex slush[23,24,25,36]) at an intermediate current drive. The other interesting observation is the absence of a vortex glass down to very low temperatures. If the pinned VL indeed survives down to the limit of T = 0, the pertinent question would be whether below a certain temperature it transforms into a quantum liquid, where quantum rather than thermal fluctuations dominate. This question is relevant to the ongoing debate on the existence of a Bose metal[40,41,42,43,44,15], and needs to be addressed in future experimental and theoretical studies.

Acknowledgements: The work was supported by Department of Atomic Energy, Govt. of India.

RD, SB and VB synthesized the sample and performed basic characterisation. RD performed STS measurements and analysed the data. SS and RT performed magnetotransport measurements and analysed the data. CD provided theoretical support. PR conceived the problem supervised the experiments and wrote the manuscript. All authors read the manuscript and commented on the paper.

[1] P. Berghuis, A. L. F. van der Slot, and P. H. Kes, Dislocation-mediated vortex-lattice melting in thin films of a-Nb$_3$Ge, Phys. Rev. Lett. 65, 2583 (1990).

[2] I. Guillamón, H. Suderow, A. Fernández-Pacheco, J. Ses´e, R. Córdoba, J. M. De Teresa, M. R. Ibarra, and S. Vieira, Direct observation of melting in a two-dimensional superconducting vortex lattice, Nat. Phys. **5**, 651 (2009).

[3] I. Guillamón, R. Córdoba, J. Ses´e, J. M. De Teresa, M. R. Ibarra, S. Vieira, and H. Suderow, Enhancement of long-range correlations in a 2D vortex lattice by an incommensurate 1D disorder potential, Nat. Phys. 10, 851 (2014).

[4] I. Roy, S. Dutta, A. N. Roy Choudhury, S. Basistha, I. Maccari, S. Mandal, J. Jesudasan, V. Bagwe, C. Castellani, L. Benfatto, and P. Raychaudhuri, Melting of the Vortex Lattice through Intermediate Hexatic Fluid in an a-MoGe Thin Film, Phys. Rev. Lett. 122, 047001 (2019).

[5] H. Pastoriza, M. F. Goffman, A. Arribere, and F. de la Cruz, First order phase transition at the irreversibility line of $Bi_2Sr_2CaCu_2O_8$, Phys. Rev. Lett. 72, 2951 (1994).

[6] E. Zeldov, D. Majer, M. Konczykowski, V. B. Geshkenbein, V. M. Vinokur and H. Shtrikman, Thermodynamic observation of first-order vortex-lattice melting transition in $Bi_2Sr_2CaCu_2O_8$, Nature 375, 373 (1995).

[7] S. S. Banerjee, S. Goldberg, A. Soibel, Y. Myasoedov, M. Rappaport, E. Zeldov, F. de la Cruz, C. J. van der Beek, M. Konczykowski, T. Tamegai, and V. M. Vinokur, Vortex Nanoliquid in High-Temperature Superconductors, Phys. Rev. Lett. 93, 097002 (2004).

[8] Lu Li, J. G. Checkelsky, Seiki Komiya, Yoichi Ando and N. P. Ong, Low-temperature vortex liquid in $La_{2-x}Sr_xCuO_4$, Nature Physics 3, 311 (2007).

[9] E. M. Forgan, S. J. Levett, P. G. Kealey, R. Cubitt, C. D. Dewhurst, and D. Fort, Intrinsic Behavior of Flux Lines in Pure Niobium near the Upper Critical Field, Phys. Rev. Lett. 88, 167003 (2002).

[10] C. J. Bowell, R. J. Lycett, M. Laver, C. D. Dewhurst, R. Cubitt, and E. M. Forgan, Absence of vortex lattice melting in a high-purity Nb superconductor, Phys. Rev. B 82, 144508 (2010).

[11] S. S. Banerjee, S. Ramakrishnan, A. K. Grover, G. Ravikumar, P. K. Mishra, V. C. Sahni, C. V. Tomy, G. Balakrishnan, D. Mck. Paul, P. L. Gammel, D. J. Bishop, E. Bucher, M. J. Higgins, and S. Bhattacharya, Peak effect, plateau effect, and fishtail anomaly: The reentrant amorphization of vortex matter in $2H\text{-}NbSe_2$, Phys. Rev. B 62, 11838 (2000).

[12] M. Zehetmayer, How the vortex lattice of a superconductor becomes disordered: A study by scanning tunneling spectroscopy, Sci. Rep. 5, 9244 (2015).

[13] S. C. Ganguli, H. Singh, G. Saraswat, R. Ganguly, V. Bagwe, P. Shirage, A. Thamizhavel, and P. Raychaudhuri, Disordering of the vortex lattice through successive destruction of positional and orientational order in a weakly pinned $Co_{0.0075}NbSe_2$ single crystal, Sci. Rep. 5, 10613 (2015).

[14] Surajit Dutta, Indranil Roy, John Jesudasan, Subir Sachdev, and Pratap Raychaudhuri, Evidence of zero-point fluctuation of vortices in a very weakly pinned a-MoGe thin film, Phys. Rev. B 103, 214512 (2021)

[15] Surajit Dutta, John Jesudasan, and Pratap Raychaudhuri, Magnetic field induced transition from a vortex liquid to Bose metal in ultrathin a-MoGe thin film, Phys. Rev. B 105, L140503 (2022)

[16] A I Larkin and Y. N. Ovchinnikov, Pinning in type II superconductors, J. Low Temp. Phys. 34, 409 (1979).

[17] M. V. Feigel'man, V. B. Geshkenbein, A. I. Larkin, and V. M. Vinokur, Theory of Collective Flux Creep, Phys. Rev. Lett. 63, 2303 (1989).

[18] Daniel S. Fisher, Matthew P. A. Fisher, and David A. Huse, Thermal fluctuations, quenched disorder, phase transitions, and transport in type-II superconductors, Phys. Rev. B 43, 130 (1991).

[19] V.M. Vinokur, P.H. Kes, A.E. Koshelev, Flux pinning and creep in very anistropic high temperature superconductors, Physica C 168, 29 (1990).

[20] C. Dekker, P. J. M. Wöltgens, R. H. Koch, B. W. Hussey, and A. Gupta, Absence of a finite-temperature vortex-glass phase transition in two-dimensional $YBa_2Cu_3O_{7-\delta}$ films, Phys. Rev. Lett. 69, 2717 (1992).

[21] V. M. Vinokur, M. V. Feigel'man, V. B. Geshkenbein, and A. I. Larkin, Resistivity of High-Tc Superconductors in a Vortex-Liquid State, Phys. Rev. Lett. 65, 259 (1990).

[22] Chandan Dasgupta and Oriol T. Valls, Phase diagram of the vortex system in layered superconductors with strong columnar pinning, Phys. Rev. B 72, 094501 (2005).

[23] T. K. Worthington, M. P. A. Fisher, D. A. Huse, John Toner, A. D. Marwick, T. Zabel, C. A. Feild, and F. Holtzberg, Observation of separate vortex-melting and vortex-glass transitions in defect-enhanced $YBa_2Cu_3O_7$ single crystals, Phys. Rev. B 46, 11854 (1992).

[24] Weiwei Zhao, Cui-Zu Chang, Xiaoxiang Xi, Kin Fai Mak and Jagadeesh S Moodera, Vortex phase transitions in monolayer FeSe film on $SrTiO_3$, 2D Mater. 3, 024006 (2016).

[25] M. Andersson, P. Fivat, L. Fàbrega, H. Obara, M. Decroux, J.-M. Triscone, and O. Fischer, ortex solid-to-liquid transition in $DyBa_2Cu_3O_{7-\delta}$ /$(Y_{0.45}Pr_{0.55})Ba_2Cu_3O_{7-\delta}$ multilayers, Phys. Rev. B 54, 675 (1996).

[26] Feng Wang, Di Zhou, Yilong Han, Melting of Colloidal Crystals, Adv. Funct. Mater. 26, 8903 (2016).

[27] Huang, P., Schönenberger, T., Cantoni, M. et al. Melting of a skyrmion lattice to a skyrmion liquid via a hexatic phase. Nat. Nanotechnol. 15, 761–767 (2020).

[28] Veatch, S., Soubias, O., Keller, S. & Gawrisch, K. Critical fluctuations in domain-forming lipid mixtures, Proc. Natl Acad. Sci. USA 45, 17650–17655 (2007).

[29] H. Suderow, I. Guillamón, J. G. Rodrigo, and S. Vieira, Imaging superconducting vortex cores and lattices with a scanning tunneling microscope, Supercond. Sci. Technol. 27, 063001 (2014).

[30] Anand Kamlapure, Garima Saraswat, Somesh Chandra Ganguli, Vivas Bagwe, Pratap Raychaudhuri, and Subash P. Pai, A 350 mK, 9 T scanning tunneling microscope for the study of superconducting thin films on insulating substrates and single crystals, Rev. of Sci. Instrum. 84, 123905 (2013).

[31] A. M. Troyanovski, J. Aarts and P. H. Kes, Collective and plastic vortex motion in superconductors at high flux densities, Nature 399, 665 (1999).

[32] Edwin Herrera, José Benito-Llorens, Udhara S. Kaluarachchi, Sergey L. Bud'ko, Paul C. Canfield, Isabel Guillamón, and Hermann Suderow, Vortex creep at very low temperatures in single crystals of the extreme type-II superconductor $Rh_9In_5S_4$, Phys. Rev. B 95, 134505 (2017).

[33] Roland Willa, Jose Augusto Galvis, Jose Benito-Llorens, Edwin Herrera, Isabel Guillamon, and Hermann Suderow, Thermal creep induced by cooling a superconducting vortex lattice, Phys. Rev. Research 2, 013125 (2020).

[34] G. I. Menon, C. Dasgupta, H. R. Krishnamurthy, T. V. Ramakrishnan, and S. Sengupta, Density-functional theory of flux-lattice melting in high-$T_c$ superconductors, Phys. Rev. B 54, 16192 (1996).

[35] Surajit Dutta, Vivas Bagwe, Gorakhnath Chaurasiya, A. Thamizhavel, Rudheer Bapat, Pratap Raychaudhuri, Sangita Bose, Superconductivity in amorphous $Re_xZr$ (x≈6) thin films, Journal of Alloys and Compounds 877, 160258 (2021).

[36] See supplementary material for (i) basic characterisation and (ii) topography of the film; (iii) filtering of the vortex lattice images; (iv) global motion path at 20 kOe; (v) STS images at 1 kOe; (vi) Non-linear vortex creep and vortex slush behaviour in the $I$-$V$ characteristics; (vii) Determination of critical current $I_c$ and (viii) Criterion for connecting $G_N(V, \boldsymbol{r})$ to obtain the red network of vortex movement.

[37] M. Buchacek, Z. L. Xiao, S. Dutta, E. Y. Andrei, P. Raychaudhuri, V. B. Geshkenbein, and G. Blatter, Experimental test of strong pinning and creep in current-voltage characteristics of type-II superconductors, Phys. Rev. B 100, 224502 (2019).

[38] C. Dasgupta and O. T. Valls, Phase diagram of vortex matter in layered high-temperature superconductors with random point pinning, Phys. Rev. B 74, 184513 (2006).

[39] A. Benyamini, E. J. Telford, D. M. Kennes, D. Wang, A. Williams, K. Watanabe, T. Taniguchi, D. Shahar, J. Hone, C. R. Dean, A. J. Millis and A. N. Pasupathy, Fragility of the dissipationless state in clean two-dimensional superconductors, Nat. Phys. 15, 947 (2019).

[40] P. Phillips and D. Dalidovich, The elusive Bose metal, Science 302, 243 (2003).

[41] J. Wu and P. Phillips, Vortex glass is a metal: Unified theory of the magnetic-field and disorder-tuned Bose metals, Phys. Rev. B 73, 214507 (2006).

[42] B. Spivak, P. Oreto, and S. A. Kivelson, Theory of quantum metal to superconductor transitions in highly conducting systems, Phys. Rev. B 77, 214523 (2008).

[43] C. Yang, Y. Liu, Y. Wang, L. Feng, Q. He, J. Sun, Y. Tang, C. Wu, J. Xiong, W. Zhang, X. Lin, H. Yao, H. Liu, G. Fernandez, J. Xu, J. M. Valles, Jr., J. Wang, and Y. Li, Intermediate bosonic metallic state in the superconductor-insulator transition, Science 366, 1505 (2019).

[44] T. Ren and A. M. Tsvelik, How magnetic field can transform a superconductor into a Bose metal, New J. Phys. 22, 103021 (2020).

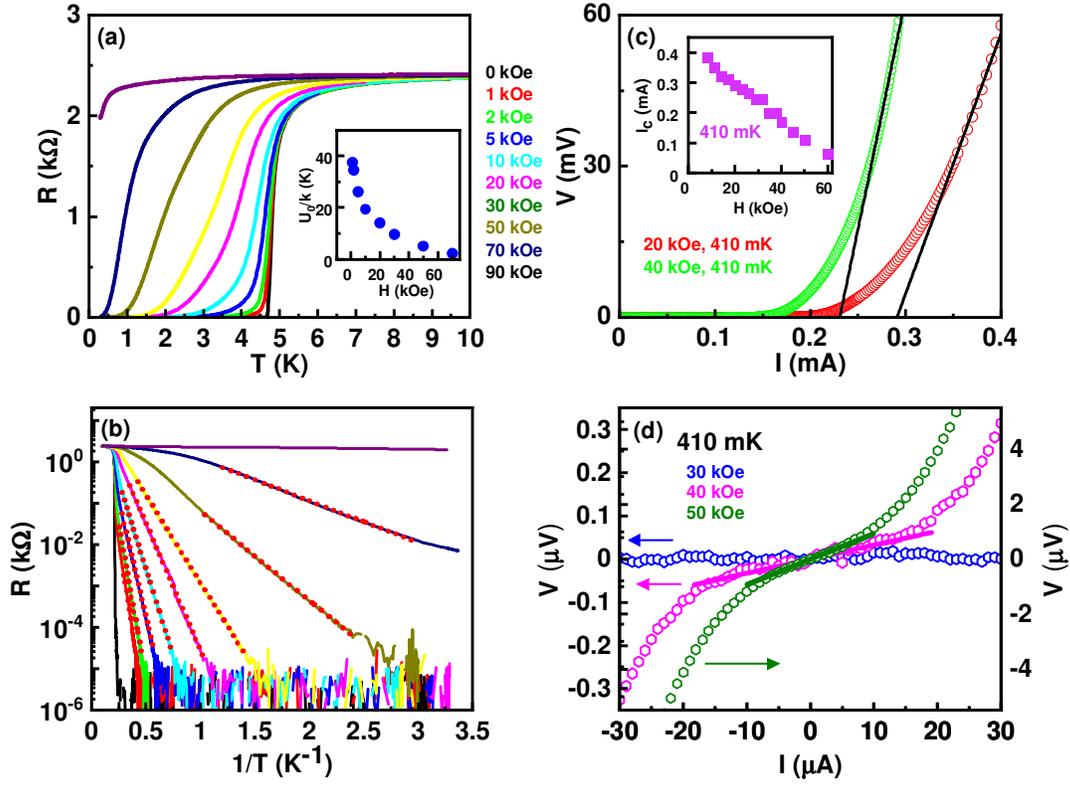

FIG.1. (a) $R$ vs $T$ and (b) $R$ vs $1/T$ (in semi log scale) in different magnetic fields ($H$); the red dotted lines in (b) are fits to the activated behaviour, $R = R_0 e^{-U(H)/kT}$. *Inset* of (a) shows $H$ dependence of activation energy *U(H)/k*. (c) Representative current-voltage ($I - V$) characteristics at $T$ = 410 mK for different magnetic fields. $I_c$ is determined from the $I$ intercept of the extrapolated linear fits (black line) to the flux flow regime. (*inset*) H dependence of $I_c$ at $T$ = 410 mK. (d) Expanded view of representative $I - V$ curves at $T$ = 410 mK for different magnetic fields. Solid lines show the linear fit to the TAFF region.

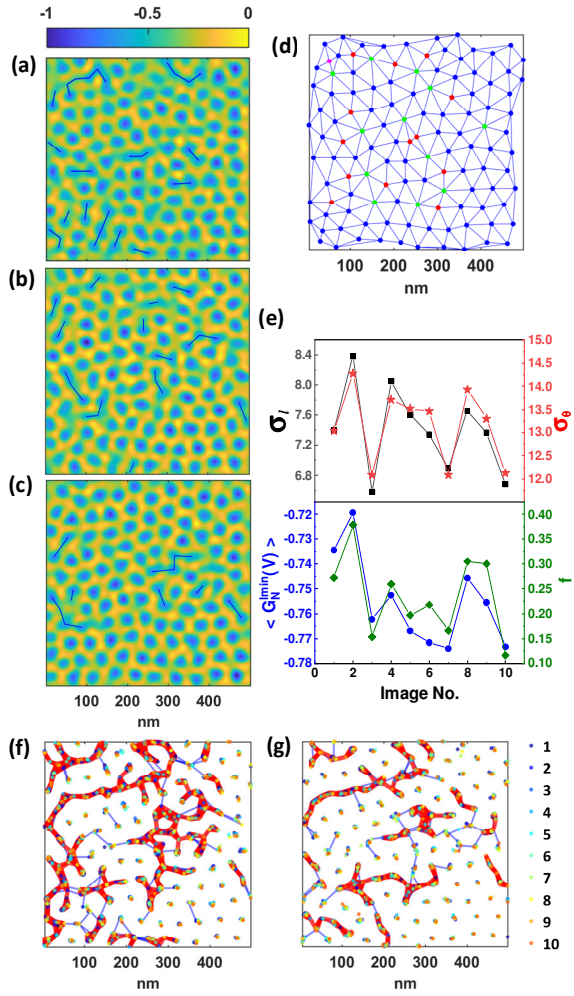

FIG. 2. (a)-(c) $G_N(V, \boldsymbol{r})$ maps of $1^{st}$,$4^{th}$,$7^{th}$ of 10 consecutive vortex images taken at 410 mK, 10 kOe. Blue lines show the motion paths in each image. (d) Delaunay Triangulation of local $G_N(V, \boldsymbol{r})$ minima (solid dots) for the image shown in (a); topological defects are color coded as magenta, red and green corresponding to four, five and seven-fold coordination respectively. (e) Variation of $\sigma_l$ (square), $\sigma_\theta$ (star), $\langle G_N^{lmin}(V) \rangle$ (circle) and $f$ (diamond) for 10 consecutive vortex images. (f) Positions of $G_N(V, \boldsymbol{r})$ minima obtained from all 10 images, where minima corresponding to each image is shown in a different colour; the red and blue networks represent the global motion paths of vortices. (g) Same as (f) but for the vortex state prepared using field cooled protocol.

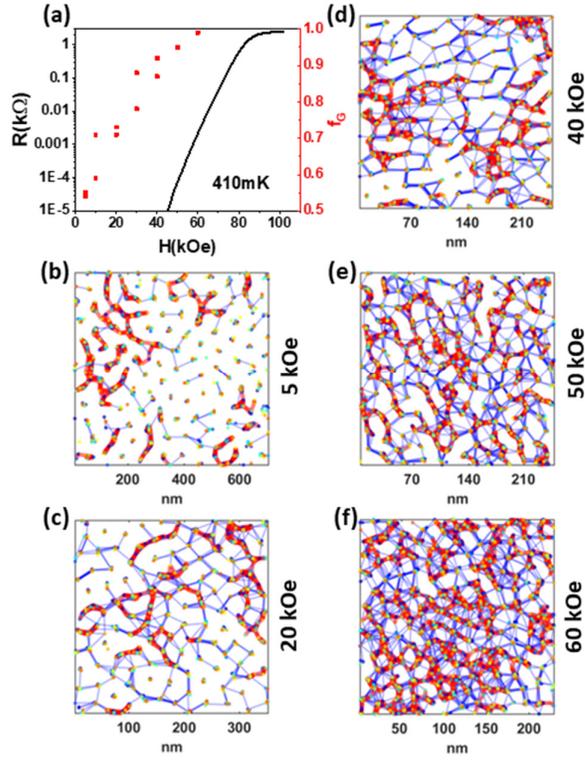

FIG. 3. (a) Variation of $R$ and $f_G$ with $H$ at 410 mK. At some fields $f_G$ was calculated from two sets of images taken at two different areas to get an estimate of statistical error. (b)-(f) Positions of $G_N(V, \boldsymbol{r})$ minima from 10 consecutive images taken at different magnetic field along with global motion paths of vortices; the colour scheme is same as Fig. 2(f).

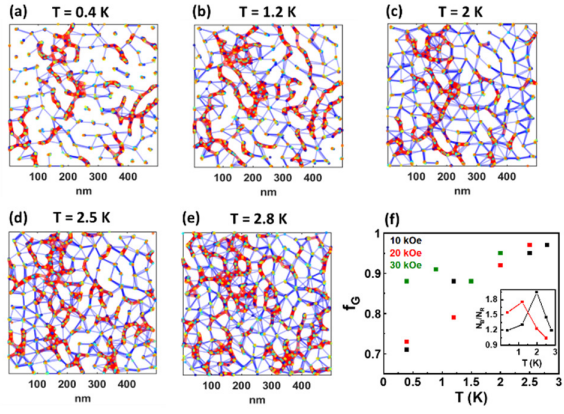

FIG. 4. (a)-(e) Positions of $G_N(V, \boldsymbol{r})$ minima from 10 consecutive images taken at 10 kOe at different temperatures along with global motion paths of vortices; the colour scheme is same as Fig. 2(f). (f) $f_G$ vs. $T$ for 10 kOe, 20 kOe and 30 kOe. (*inset*) $\left(\frac{N_B}{N_R}\right)$ variation with temperature for 10 and 20 kOe respectively.

## Supplementary Material: Structure and dynamics of a pinned vortex liquid in superconducting $a$-Re$_x$Zr (x ≈ 6) thin film

Rishabh Duhan[1], Subhamita Sengupta[1], Ruchi Tomar[1], Somak Basistha[1], Vivas Bagwe[1], Chandan Dasgupta[2] and Pratap Raychaudhuri[11]

[1] *Tata Institute of Fundamental Research, Homi Bhabha Road, Colaba, Mumbai 400005, India.*
[2] *Department of Physics, Indian Institute of Science, Bangalore 560012, India.*

### I. Basic characterisation of superconducting properties:

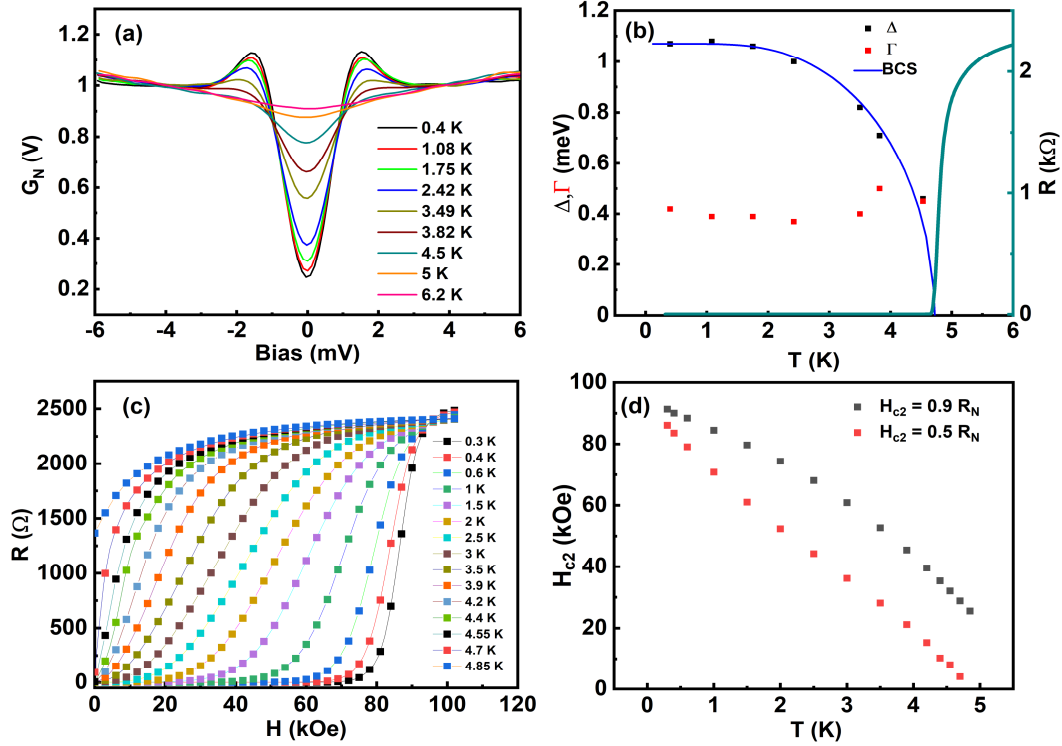

FIG. 1S. (a) Normalised Tunnelling Spectra $G_N(V)$ vs. applied bias at different temperature. (b) Variation of $\Delta, \Gamma$ and resistance as function of temperature. Blue line is expected variation of $\Delta(T)$ obtained from BCS theory. (c) $R$ vs. $H$ plotted at different temperature. (d) Variation of $H_{c2}$ with temperature.

The superconducting energy gap ($\Delta$) is determined by acquiring zero field tunnelling spectra at different temperature using STM (Fig. 1S(a)). These spectra were then fitted using normalised tunnelling conductance $G_N(V) = \frac{1}{R_N} \int N_S(E) \frac{-\delta f(E - eV)}{\delta(E)} dE$, where $f(E)$ is Fermi Dirac distribution function, $R_N$ is tunnelling resistance at high bias and $N_S(E) = Re\left(\frac{|E| + i\Gamma}{\sqrt{(|E| + i\Gamma)^2 - \Delta^2}}\right)$ is DOS of superconducting sample with $\Gamma$ as phenomenological parameter taking account of non-thermal broadening of spectra. We obtain $\Delta(0) = 1.08\ meV$ with $\Delta(T)$ following BCS type variation. Resistance appears as superconducting gap approaches zero near $T_c$ (Fig. 1S (b)).

In Fig. 1S(c) we show the *R-H* plots at various temperatures. In Fig. 1S(d) we show the estimated $H_{c2}$ using two different criteria: The field at which the resistance is 90% of the normal state value and the field where the resistance is 50% of the normal state value.

## II. Topography of sample:

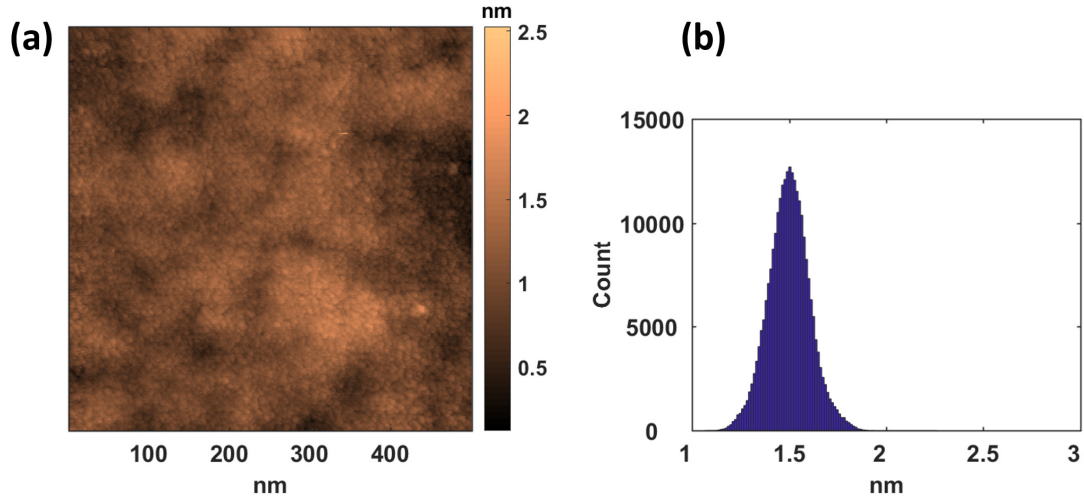

FIG 2. (a) Representative topographic image of sample surface recorded at 4 K over an area of 500 nm x 500 nm. (b) Histogram of surface height.

The topography of the sample was characterised at 4 K using STM. Measurements were performed at a bias voltage of 10 mV using a tunnelling current of 260 pA. As can be seen from Fig. 2S, the surface is smooth with very few particulates. We find rms roughness of about 0.21 nm.

## III. Filtering $G(V, r)$ maps:

$G(V, r)$ maps taken using STS were filtered to remove the noise in the image. Filtering is done as per process shown in Fig. 3S. Figure 3S(a) shows the raw image. We first perform 2D Fast Fourier Transform (FFT) of this image (Fig. 3S(b)). The FFT shows a bright ring corresponding to the disordered vortex lattice, along with a diffuse background at higher q-values corresponding to high frequency noise. (Fig. 3S(b)). To remove the noise, we suppress this intensity at high q values as shown in Fig. 3S(c) and obtain the filtered image by taking an inverse FFT. The filtered image is shown in Fig. 3S(d).

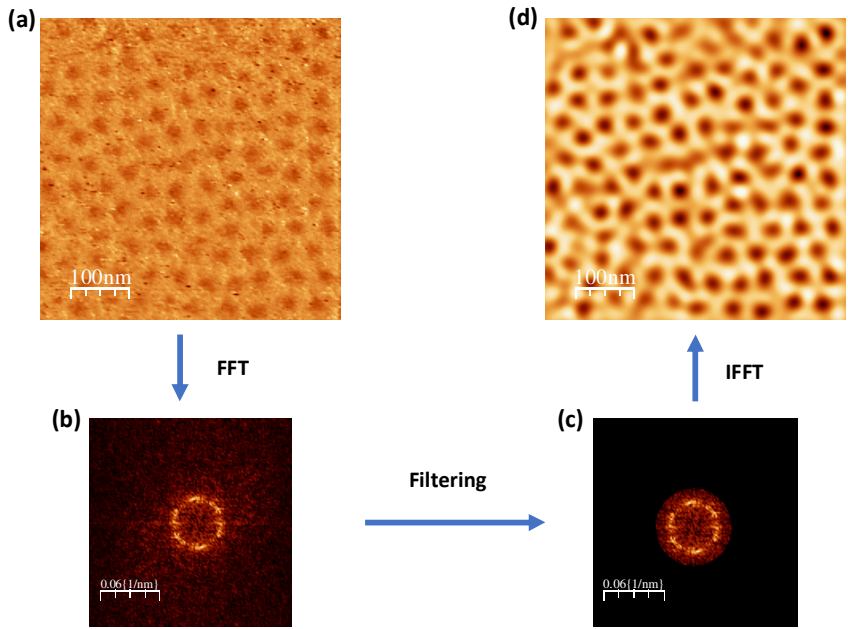

FIG. 3S. (a) Raw image recorded from STM. (b) FFT of image. (c) Filtered FFT having noise removed at high frequency. (d) Filtered image by taking inverse FFT of (c).

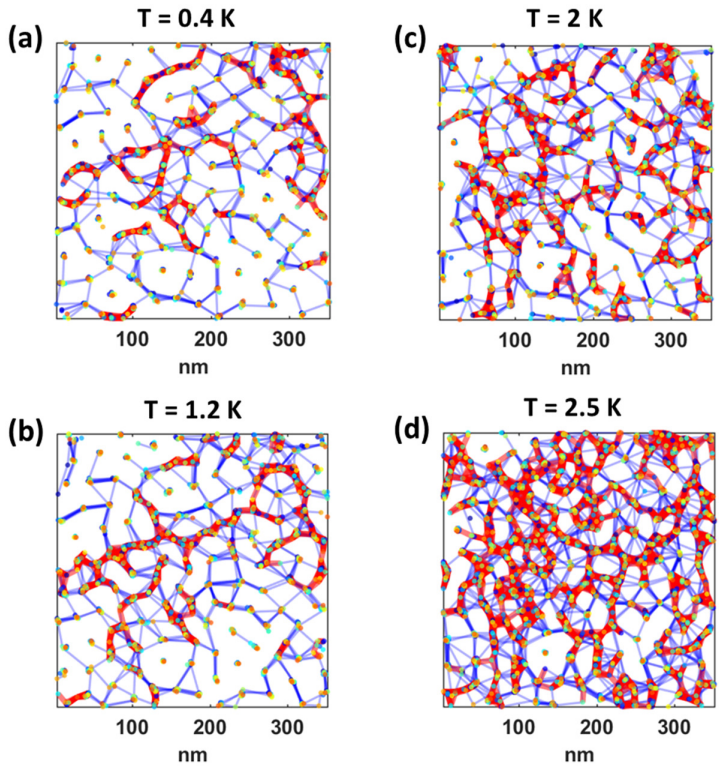

FIG. 4S. (a)-(d) Positions of $G_N(V, \boldsymbol{r})$ minima from 10 consecutive images taken at 20 kOe for different temperatures along with global motion paths of vortices.

## IV. Global vortex movement network at 20 kOe:

The evolution of red and blue line network (as described in main text) with various temperature for images taken at 20 kOe is shown in Fig. 4S. With increase in temperature gradually the fraction of vortices taking part in motion increases till this network spreads over the entire area.

## V. Vortex Liquid at 1 kOe:

Here we show STS data of the vortex lattice in very low field of 1 kOe at 410 mK. At this field the density of vortices is low and the imaging had to be performed over a larger area of $1.3 \times 1.3$ μm. Consequently, we recorded only 5 consecutive images in the same area. Fig. 5S (a)-(c) show the $1^{st}$, $3^{rd}$ and $5^{th}$ $G_N(V, \boldsymbol{r})$ maps in the sequence. Here all minima are well separated from each other and complete hops of vortices between two minima are extremely rare owing to large inter-vortex distance and stronger interaction between vortices and defect pinning centres. However, the motion of vortices becomes apparent when we track the positions of minima from the 5 consecutive images. Fig. 5S(d) shows the positions of minima obtained from $G_N(V,r)$ maps from all 5 images along with the red network defining the motion paths as described in the main text.

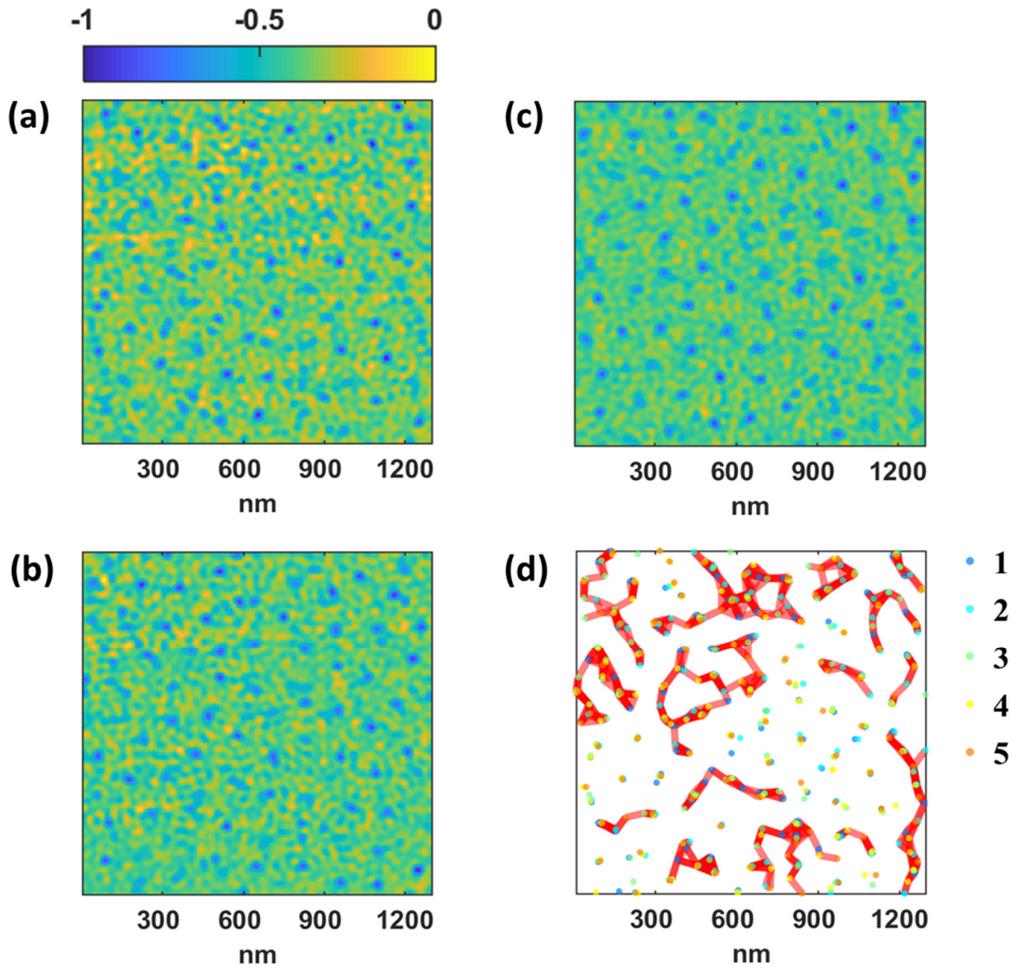

FIG. 5S. (a)-(c) $G_N(V, \boldsymbol{r})$ maps for $1^{st}$, $3^{rd}$, $5^{th}$ images respectively taken at 410 mK in magnetic field of 1 kOe. (d) Positions of $G_N(V, \boldsymbol{r})$ minima for 5 consecutive images along with global motion paths of vortices.

## VI. Very low onset of non-linear creep and vortex slush behaviour:

Superconducting thin films have been known to exhibit several unusual I-V characteristics in the mixed state. For example, it has been reported that the non-linear creep sets in a very low currents and the linear TAFF region keeps shrinking rapidly with decrease in temperature. Here we determine the onset of nonlinear creep in our film using the same criterion as used in ref. 20, i.e. the current $I_{nl}$ at which $\frac{\partial ln(V)}{\partial ln(I)} = 1.2$ (Fig. 6S(a)). In Fig. 6S(b) we plot $I_{nl}$ as a function of temperature for 20 and 30 kOe. $I_{nl}$ drops rapidly with decreasing temperature tends towards zero at T = 0 in a non-linear fashion. The other observation reported in several samples is signatures of vortex slush where the resistance as a function of current has an S-shape, and appears to plateau at a nearly temperature independent value much lower than the expected Bardeen-Stephen flux flow resistance at an intermediate current drive. In our sample we observe signature of vortex slush behaviour below 1.4 K (Fig. 6S(a)).

We believe that both these behaviours can be understood from the percolative transport network that we observe in STS studies. Just like with increase in temperature or magnetic field, with increase in current the percolative network is expected to get denser such that the number of vortices participating in dissipation will increase continuously with applied current. This would give a faster than linear increase the voltage in at low currents. Within this picture, the plateau is expected when the network connects all vortices. At this drive all vortices are mobile with their velocities having a large distribution and the average mobility will be smaller than that expected for Bardeen Stephen flux flow due to the underlying pinning potential. As the current is further increased the resistance gradually crosses over to the flux flow regime.

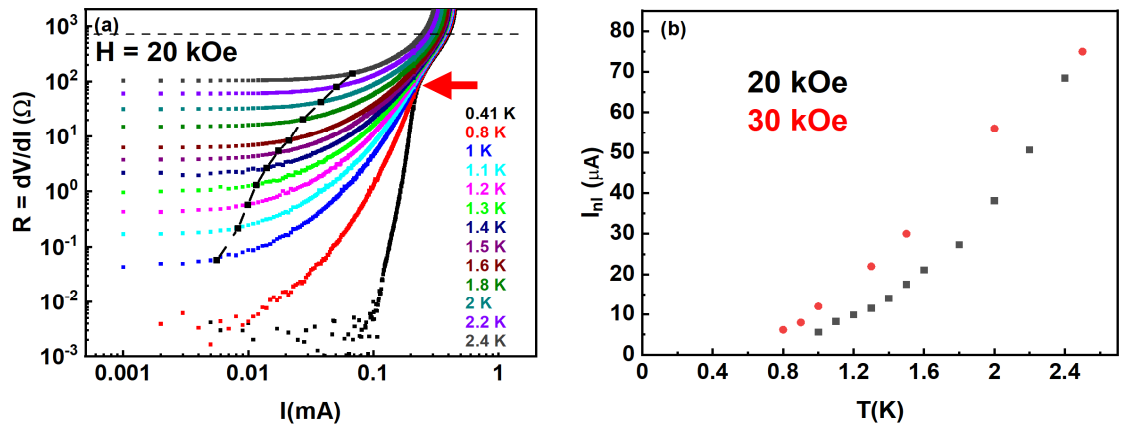

FIG. 6S. (a) $R$ vs. $I$ plotted in log-log scale for various temperature at 20 kOe. The black squares show the onset of non-linear flux creep on the $R$-$I$ curves; the connecting dashed line is a guide to the eye. The red arrow points to the vortex slush behaviour. The horizontal dashed line shown the value of $R_{ff}$ at 0.41 K. (b) Variation of $I_{nl}$ with temperature for 20 and 30 kOe.

## VII. Determination of Bardeen-Stephen critical current

In an idealised situation at T=0, the *I-V* curve in the mixed state is given by $V = R_{ff}(I - I_c)\Theta(I - I_c)$, where $R_{ff}$ is the Bardeen-Stephen flux flow resistance given by $R_{ff} = R_N\left(\frac{H}{H_{c2}}\right)$ (where $R_N$ is the normal state resistance). At finite temperatures vortex creep causes the sharp cut-off at $I_c$ to be rounded off. It is therefore customary to extrapolate the linear flux flow region of the *I-V* curve to obtain the value of $I_c$ from the intercept with the current axis.

In 2-dimensional films, the formation of the vortex slush complicates the scenario. Since the flux flow regime is gradually reached through a continuous increase in resistance after the onset of vortex slush, the linear flux flow regions in the *I-V* curve shrinks to a very small range of current, before the onset of depairing that drives the superconductor to the normal state. Here, to estimate the $I_c$ we first calculate $R_{ff}$ from the normal state resistance, $R_N$, and upper critical field and draw a tangent to the *I-V* curve where the slope matches $R_{ff}$. $I_c$ is taken where the tangent intercepts the current axis (Fig. 7S).

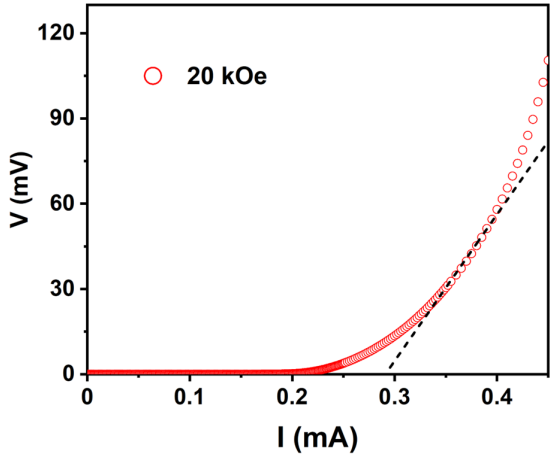

Fig. 7S. *I-V* curve at 410 mK, 20 kOe. The dashed line is the tangent to the regions of the curve where the slope is $R_{ff}$. $I_c$ is obtained from the intercept of the dashed line with the *I* axis.

## VIII. Criterion for connecting $G_N(V, r)$ minima to obtain the red network of vortex movement

To obtain the global network from incomplete hops (the red networks in the paper), we connected the nearby $G_N(V, r)$ minima obtained in different images acquired at the same area that are separated by a distance, $d \leq 0.5a_H$. In this network we also excluded isolated segments smaller than $a_H$ which correspond to meandering motion of a vortex about its mean position. The criterion for $d$ has to be carefully chosen: Using a smaller cut-off results in missing out on many path, whereas a larger cut-off results in over-connecting minima.

In Fig. 8S, we illustrate how we chose this criterion. In Fig. 8S(a) we plot the sum of a sequence of 10 images, $G_N^{SUM}(V, \boldsymbol{r}) = \sum_{i=1}^{10} \boldsymbol{G}_N^i(V, \boldsymbol{r})$ (where $i$ is the index of the image in the sequence), taken at 10 kOe, 410 mK. On this map the conductance at the minima that that remain at the same location in each individual image add up to give deep minima where $G_N^{SUM}(V, \boldsymbol{r}) < -8$. In the other hand the minima that appear at different locations in each image form shallow tracks where $G_N^{SUM}(V, \boldsymbol{r}) \gtrsim -6$. These tracks define the global motion path of vortices from incomplete hops. In Fig. 8S(b)-(f) we now show the red network by using different cut-offs. We clearly see that using a cut-off of less than $0.5a_H$ misses many connections in the network. On the other hand, a cut-off greater than $0.5a_H$ over-connects points that are deep minima, and therefore not mobile. The trade-off seems to be around $0.5a_H$ which captures the network formed by shallow minima most faithfully. This was verified by inspecting several images at different fields and temperatures.

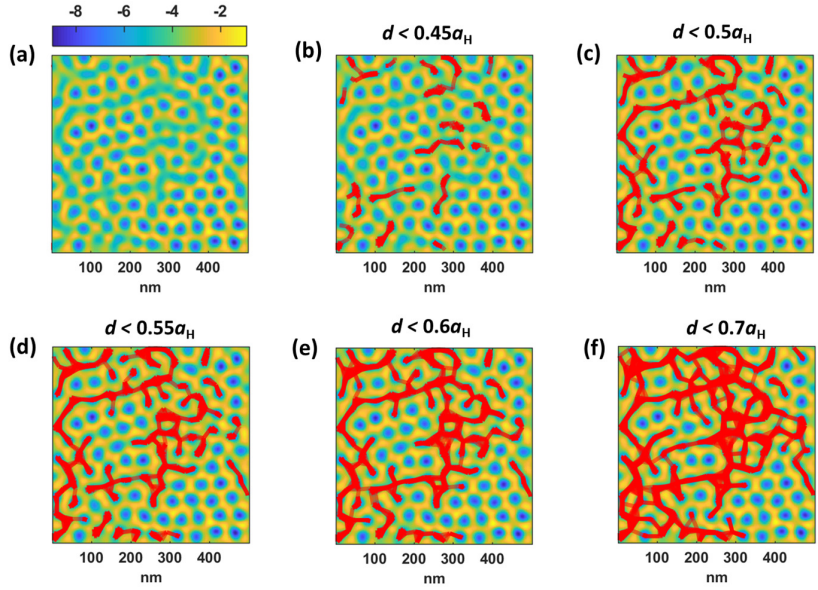

Fig. 8S. (a) $G_N^{SUM}(V = 1.5\ mV, \boldsymbol{r})$ map at 410 mK, 10 kOe. (b)-(f) red network superposed on $G_N^{SUM}(V, \boldsymbol{r})$ image calculated by using different criteria for joining minima.